# 4[th] Generation Light Sources and elementary particle physics

E.G.Bessonov

Lebedev Phys. Inst. RAS, Moscow, Russia

In this review a history of development of electron accelerators for HEP and for LSs in Russia is presented.

## Introduction

"Synchrotron radiation was first observed in 1947 and began to be used as a research tool in the mid-1960s. The first synchrotron radiation sources (**1[st] generation**) were low energy (several hundred MeV) **electron accelerators which were first of all meant to be used for nuclear and particle physics.** It was soon realized that the needs of the growing number of people who used synchrotron radiation for their research required **dedicated sources of radiation.** Many of these research programs required X-radiation which pointed to the need for these sources to accelerate electrons to GeV energies in order to provide a useful output of X-rays as well as lower energy radiation. **The first such a dedicated X-ray source (2[nd] generation) was the SRS (Synchrotron Radiation Source) at the Daresbury Laboratory in the UK**, which operated at 2GeV electron energy. In the early days the emphasis in research applications was on the exploitation of VUV radiation for atomic and molecular spectroscopy and surface science but the advent of the X-ray sources stimulated a steady growth of research with harder radiation in such areas as X-ray diffraction, absorption spectroscopy (EXAFS), crystallography, topography, and lithography. **The users of the radiation, initially drawn from the low and high energy physics (HEP) community, have been augmented by lower (X-ray) energy physics community, chemists, biologists, geologists, and others** so that there is now a world-wide community of people using synchrotron radiation as an essential component of their research" (taken from [1]).

"Synchrotron and undulator radiation has become the brilliant source of photons from the infrared to hard X-rays for a large community of research in basic and applied sciences. **This process was particularly supported by the development of electron accelerators for basic research in high energy physics.** Specifically, the development of the storage ring and associated technologies resulted in the availability of high brightness photon beams far exceeding other sources" (taken from [2]).

At present all sources based on relativistic electron and ion beams are named Light Sources (LS). Up to now three Generations of LSs were developed. The brilliance and hardness of sources was increased from one generation to another by developing more and more perfect magnetic lattices of the rings to reach smallest emittances of the stored beams. At present the 4[th] Generation LSs (4GLS) are developed in different scientific centers. The idea of 4GLSs is based on the possibility to obtain high energy, high brilliance, high current, continuous electron beams in superconducting energy recovery linear accelerators. It was suggested by M.Tigner in 1965 and intended first of all to linear colliders [3]. He noticed as well that the energy recovery technique might be useful in experiments other than clashing (colliding) beam type (e.g., a low-density target such as liquid hydrogen might be placed in the return leg of the magnet system for nuclear physics). Now the idea of 7 GeV 4GLS based on the Energy Recovered Linac (ERL) is developed in Cornell [4] (Fig. 1). More compact race track, polygon microtron and other types of energy recovery recirculators are considered for the 4GLSs [5-7] (Fig. 2-5). CEBAF type 4GLSs (Fig. 6) with energy recovery and additional straight sections for insertion devices can be used. The idea of the ERL was tested [8] (see Fig. 5). Energy recovery linacs have made great strides in the past decade and are now poised to evolutionize light sources, lepton-hadron colliders, electron coolers, high-power free electron lasers (FELs), Compton sources, THz radiators [9].

In this review a history of development of electron accelerators for HEP and for LSs in Russia is presented. The using of 4GLSs simultaneously or in turn both in low energy scientific disciplines and in HEP is discussed. Such a way low energy community can be augmented by HEP community this complicated time.

# From accelerators and storage rings for high energy physics to 4GLS and back to using 4GLSs in high energy physics

The development of the nuclear and elementary particle physics for the last half of the previous century was determined by the development of particle accelerators and storage rings. In time these facilities became to be used in other fields of science and technology. At present the progress in these fields is closely connected with the development and use of LSs based on relativistic electron beams in storage rings. One of the main motivations to build such sources is to have a very brilliant source of monochromatic photon beams with smoothly varied frequency in a wide spectral region and different kinds of polarization. This might be achieved by production of low emittance electron beams in accelerators and storage rings and the use of a variety of undulators that can be optimized to the special demands of a certain experiment [10, 11]. Recently new electron synchrotrons for HEP using beam-fixed targets are not constructed anywhere (from early 70$^{th}$, see Fig. 7). That is why **the development of inter-mediate and HEP is possible in a reverse direction i.e. through development of LSs if the needs of the HEP users will be taken into account. The users of the VUV and X-radiation of 4GLSs drawn from the low energy physics community (chemistry, biology, geology, medicine, medical isotope production and others) can be augmented by high energy physics community.** Many researches remain interesting with electron and gamma beams interacting with fixed targets in HEP. Bremstrahlung radiation in the form of $\gamma$**-ray beams from beam-fixed targets**, electron beams (energy recovering regime is switched off) and backward Compton scattering (BCS) radiation (energy recovering regime is switched on) can be used in the last case. Twice repeated regime of acceleration in ERL (see Fig. 1) allows producing double energy high average power electron beam in the low current regime. Below the problem of the intermediate energy physics and 4GLSs in Russia is discussed.

# 4GLS and intermediate energy physics in Russia

Historically High Energy Physics (HEP) based on particle accelerators appeared in Lebedev Physical Institute (LPI), Moscow after V.I.Veksler invented the phase stability principle (1943). Two electron synchrotrons were constructed with the energy 30 and 250 MeV in LPI under his leadership. Then 1.2 GeV synchrotron "Pakhra" was constructed in 70$^{th}$. It is under operation now at the HEP department LPI in Troitsk near Moscow. $\eta$-nuclear and electromagnetic interaction research was developed recently at the synchrotron. In due time, the development of the cascade system of accelerators and storage rings for colliding beams at the intermediate energy region was planned [12]. Higher energy regions were considered as well. To that time first experiments with storage of colliding $e^{\pm}$ beams were produced at LPI [13] [14]. Moreover the theory and experimental technique for LSs was developed. First experience with the beam of spontaneous undulator radiation (UR) emitted from the orbit of the circular machine was done at the undulator beam line of synchrotron "Pakhra" [15]. Later, a spontaneous coherent UR source with cavity (recently named stimulated superradiant emission in the prebunched FEL) based on the microtron was produced [16]. The scheme of colliding beams was successively used at CEA [17]-[19]. Another centers constructed storage rings with colliding beams later. For these reasons the project [12] of the cascade system was not realized. Nevertheless the upgrade of the synchrotron "Pakhra" was required for a long time. More effective recirculators were developed in 80$^{th}$. Some activity in this direction was at LPI as well [20], [21]. LSs and other applications of recirculators with low emittance electron beams were discussed before [22]. Linear accelerators had low emittance (less than in storage rings). The problem of beam emittance degradation in recirculators at high energies was searched [21], [23] (energy recovering regime of M.Tigner seemed not realistic that time but very attractive). Emittance degradation can be essential and must be taken into account in the design of the lattice of the ERLs both for 4GLSs and for the HEP.

The use of 4GLSs simultaneously or in turn both in low energy scientific disciplines and in elementary particle physics has many economic advantages. In such a way the development of 4GLSs can support the existing nuclear and elementary particle physics programs. Small scale model of the source can be used as a source of gamma- and X-rays as well.

Energy recovery Racetrack microtron [5], MAMIc or CEBAF type recirculator of the energy ~1.5 - 2GeV can be installed in the hall of the synchrotron "Pakhra" on the territory of the LPI branch near Moscow (Troitsk). Upgrade is possible to the energy 5-7 GeV on the territory LPI or neighboring Institute of Nuclear Research RAS. Note that the main part of users is concentrated in Moscow and Moscow region. $3^d$ GLS SIBERIA-2 in Moscow satisfies the requirements of low energy users now [24]. ERL "Mars" was suggested for the Novosibirsk region of Russia in addition to their LSs based on storage rings [25].

## Conclusion

LS's are sources of powerful beams of IR to $\gamma$-rays having high degree directionality, narrow bandwidth, tunability, variable photon energy and polarization. These sources are based on accelerators and storage rings and make possible basic and applied research in different fields that are not possible with more conventional equipment. They are UR sources, including Backward (or Inverse) Compton scattering (BCS) sources and possibly future Backward Rayleigh scattering (BRS) sources. These sources are both spontaneous incoherent, spontaneous coherent (prebunched FEL's) and stimulated (FEL's) UR sources [11]. 4GLS will allow studying unique problems of nature and technology. Moreover the radiofrequency superconductor technique will be developed in Russia.

This paper reflects long-term discussions about the development of the electron accelerator technique for the intermediate energy elementary particle physics and physics on LSs in different centers in Russia. At present such discussions were renewed [26], [27]. 4GLS facility can be constructed as a national Project for Russia. Currently a unification of HEP and low energy physics in one LS facility is the best way to satisfy the demands of different communities in Russia.

I thank A.I.Lvov for useful remarks. Work supported by grant RFBR 09-02-00638-a.

## References

1) Philip John Duke, Synchrotron Radiation Production and Properties, OXFORD University Press, 2000.
2) H.Wiedenann, Synchrotron Radiation, Springer-Verlag Berlin Heidelberg, 2003.
3) M.Tigner, A possible apparatus for electron clashing-beam experiments, Nouvo Cimento v.37, p. 1228-1231, 1965.
4) G.H. Hoffstaetter, I.V. Bazarov, D.H. Bilderback, J. Codner, B. Dunham, D. Dale,
K. Finkelstein, M. Forster, S. Greenwald, S.M. Gruner, Y. Li, M. Liepe, C. Mayes, D. Sagan, C.K. Sinclair, C. Song, A. Temnykh, M. Tigner, Y. Xie, **PROGRESS TOWARD AN ERL EXTENSION TO CESR,** Proceedings of PAC07, Albuquerque, New Mexico, USA, p.107-109; G.H. Hoffstaetter et all, **CHALLENGES FOR BEAMS IN AN ERL EXTENSION TO CESR,** Proceedings of EPAC08, Genoa, Italy, p.190-192.
5) Mikael Eriksson, Lars-Johan Lindgren, Erik Wallen, Sverker Werin, A cascaded optical klystron on an energy recovery linac – racetrack microtron, Nuclear Instruments and Methods in Physics Research A 507 (2003) 470–474.
6) M. E. Couprie 1, M. Desmons, B. Gilquin1, D. Garzella1, M. Jablonka, A. Loulergue, J. R. Marquès, J. M. Ortega F. Méot, P. Monot1, A. Mosnier, L. Nahon1, A. Rousse, **"ARC-EN-CIEL" A PROPOSAL FOR A 4TH GENERATION LIGHT SOURCE IN FRANCE** Proceedings of EPAC 2004, Lucerne, Switzerland.
7) M W Poole, S L Bennett, M A Bowler, N Bliss, J A Clarke, D M Dykes, R C Farrow,
C Gerth, D J Holder, M A MacDonald, B Muratori, H L Owen, F M Quinn, E. A. Seddon, S L Smith, V P Suller and N R Thompson, I N Ross, B McNeil, **4GLS: A NEW TYPE OF**

**FOURTH GENERATION LIGHT SOURCE FACILITY,** Proceedings of the 2003 Particle Accelerator Conference.
8) S. Benson, D. Douglas, M. Shinn, K. Beard, C. Behre, G. Biallas, J. Boyce, H. F. Dylla, R. Evans, A. Grippo, J. Gubeli, D. Hardy, C. Hernandez-Garcia, K. Jordan, L. Merminga, G. R. Neil, J. Preble, T. Siggins, R. Walker, G. P. Williams, B. Yunn, S. Zhang, H. Toyokawa, **HIGH POWER LASING IN THE IR UPGRADE FEL AT JEFFERSON LAB,** Proceedings of the 2004 FEL Conference, 229-232.
9) L. Merminga, (Jefferson Lab, Newport News, Virginia), ENERGY RECOVERY LINACS, Abstracts of the PAC07 Conference, Albuquerque, New Mexico, USA, MOYKI03; Proc. RUPAC 2006, Novosibirsk, Russia, p. 10-12.
10) E.G.Bessonov, Electromagnetic Radiation Sources Based on Relativistic Electron and Ion Beams, J. "Radiation Physics and Chemistry" 75 (2006), p. 908-912.
11) E.G.Bessonov, Light sources based on relativistic electron and ion beams, Proc. of SPIE Vol. 6634, 66340X-1 – 66340X-14, (2007).
12) Belovintsev K.A., Belyak A.Ya., Bessonov E.G., Kolomensky A.A., Lebedev A.N., Fateev A.P., Cherenkov P.A., Cascade storage system, Proc. 6th Int. Accelerator Conf., Cambridge, 1967, USA, p.23.
13) Ado Yu.M., Belovintsev K.A. Belyak A.Ya., Bessonov E.G., Demyanovsky O.B., Skorik V.A., Cherenkov P.A., Schirchenko V.S. Storage of particles in the synchrotron, Proc. 4th Int. Accelerator Conf., Dubna, USSR, 1963, p.355.
14) Yu.M.Ado, K.A.Belovintsev, E.G.Bessonov, P.A.Cherenkov, V.S.Schirchenko, Storage of positrons and production of the colliding electron-positron beams in the synchrotron, Atomic energy, v.23, No1, 1967.
15) D.F.Alferov, Yu.A.Bashmakov, K.A.Belovintsev, E.G.Bessonov, P.A.Cherenkov, Observation of undulating radiation with the "Pakhra" synchrotron, Phys. - JETP Lett., 1977, v.26, N7, p.385-388; The Undulator as a source of electromagnetic radiation, Particle accelerators, 1979, v.9, No 4, p.223-235.
16) V.I. Alexeev, K.A.Belovintsev, E.G.Bessonov, A.V.Serov, P.A.Cherenkov, A parametric free-electron laser based on the microtron, Nucl. Instr. Meth., 1989, A282, p.436-438; see also Yukio Shibata, Kimihiro Ishi, Shuichi Ono et al., Broadband Free Electron Laser by the Use of Prebunched Electron Beam, Phys. Rev. Lett., 1997, v.78, No 14, pp. 2740-2743.
17) K.W.Robinson, T.R.Sherwood, G.A.Voss, A PROPOSED SYSTEM FOR MULTI-CYCLE INJECTION OF POSITRONS AND ELECTRONS INTO THE 6-GeV CAMBRIDGE ELECTRON ACCELERATOR, IEEE TRANSACTIONS ON NUCLEAR SCIENCE, JUNE 1967, p.670-676.
18) R. Averill, W. F. Colby, T. S. Dickinson, A. Hofmann, R. Little, 8. J. Maddox, H. Mieras, J. M. Paterson, K. Strauch, G.-A. Voss, H. Winick, "PERFORMANCE OF THE CEA AS AN e+e- STORAGE RING", 1973 IEEE, p.813-815.
19) G.A. Voss, A Personal Perspective of High Energy Accelerators, 1996 IEEE, p.27-29.
20) K.A.Belovintsev, A.I.Karev, V.G.Kurakin, Perspectives of usage of segment magnets in high energy electron recirculators, Sov. Phys. Tech. Phys. Letters, v. 10, No 7, p.439-443, 1984 (in Russian).
21) E.G.Bessonov, Effect of quantum fluctuations of synchrotron radiation on the dynamics of particles in high-energy microtrons, Sov. Phys. Tech. Phys. V 32 (5), May 1987, p. 605-607.
22) Kaiser H.P., Microtron – device for research in the fields of nuclear phisics and electronics, IEE Trans. NS 2, No 3, p. 17 - 27, 1956.
23) Crosbie E.A. IEEE Trans. Nucl. Sci, 1983, v.30, No 4, p.3218-3220.
24) V.V.Anashin, A.G.Valentinov, V.G.Veshcherevich et al., The dedicated synchrotron radiation source SIBERIA-2, Nucl. Instr. Meth., A282, !989, p.369-374.
25) D.A.Kayran, V.N.Korchuganov, G.N.Kulipanov, E.B.Levichev, V.V.Sajaev, A.N.Skrinsky, P.D.Vobly, and N.A.Vinokurov, **MARS - A PROJECT OF THE DIFFRACTION LIMITED FOURTH GENERATION X-RAY SOURCE, Proceedings of APAC98; G.N.**Kulipanov,

A.N.Srinsky, N.A.Vinokurov (1998). Synchrotron light sources and recent developments of accelerator technology. *J. Synchrotron Rad*. **5** : 176.
26) V.G.Kurakin, G.A.Sokol, V.G.Nedoresov, B.S.Ishkhanov, V.I.Shvedunov, **SUPERCONDUCTING RF ELECTRON RECIRCULATOR FOR NUCLEAR PHYSICS RESEARCH AT LEBEDEV PHYSICAL INSTITUTE** Proceedings of RuPAC 2008, Zvenigorod, Russia.
27) E.G.Bessonov, Possible ways of development of HEP department of LPI, The proceedings of the HEP department of LPI. Troitsk, 16-17 October 2008 (in print).

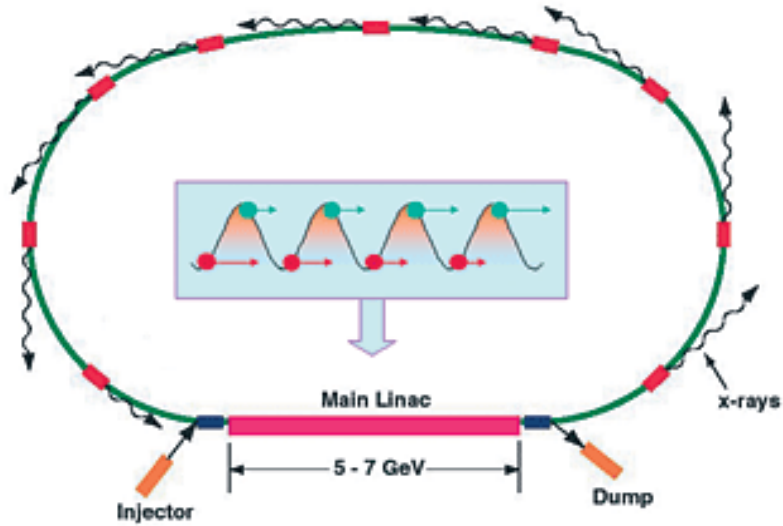

**Figure 1a. The idea of the ERL. Electrons are pushed to almost the speed of light, "surfing" on the crest of microwaves in a linear accelerator (linac). The accelerated electrons make one trip around the Cornell Electron Storage Ring (CESR), and returning electrons feed back into the linac and give up their remaining energy to the microwaves.**

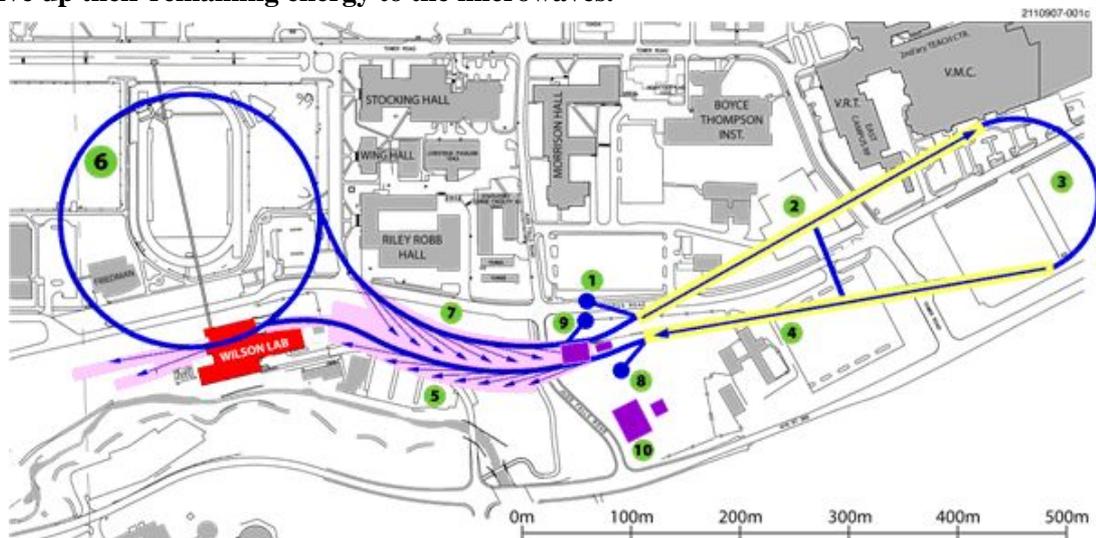

**Figure 1b: Schematic ERL layout incorporating the existing Cornell Electron Storage Ring (CESR). Electrons are injected (1) and are accelerated to the right a 2.5 GeV linac (2), loop through a turn-around arc (3), and accelerate to the left through an additional 2.5 GeV linac (4) to 5 GeV, total. Beamlines are in the pink/red areas. Bunches then pass clock-wise around CESR (6). Bunches may be compressed to <100 fs (7) and feed more undulators before being uncompressed, energy recovered in second passes though linacs (2) and (4), and finally dumped at (8). A second injector (9) provides large bunches for pump-probe experiments (mode c of Table B-1). These are accelerated to 2.5 GeV through linac (4), pass through a long undulator,and are dumped without energy recovery at (8). The pump-probe station is at (10).**

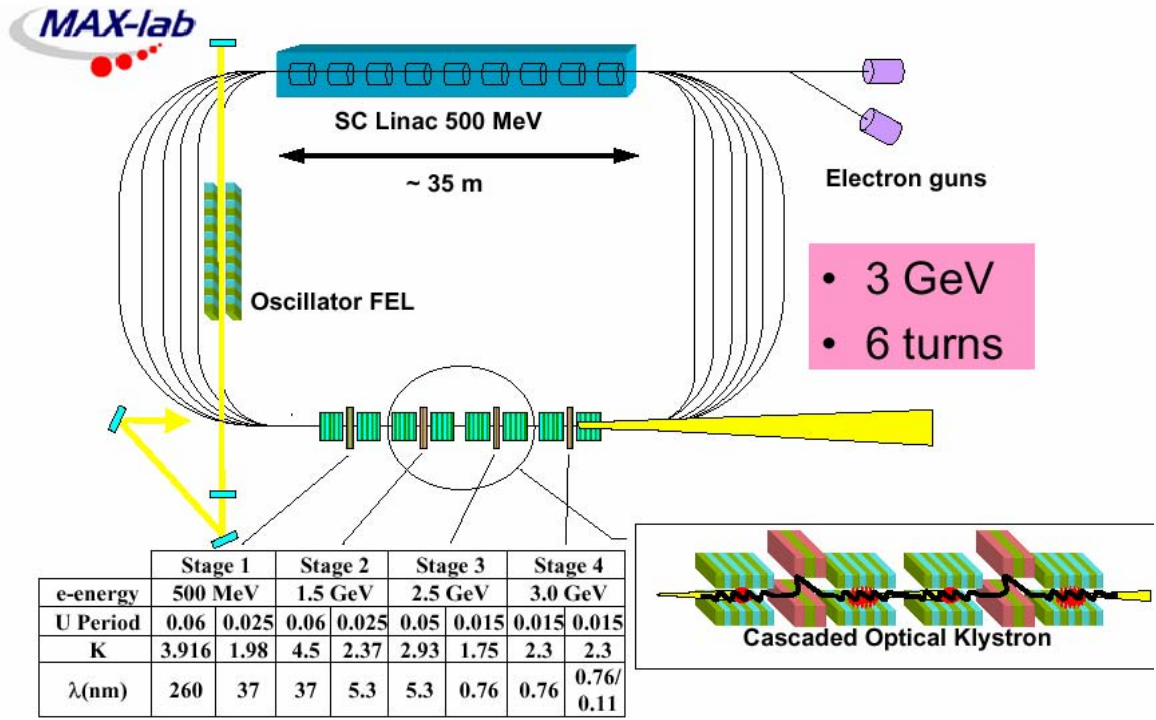

Figure 2

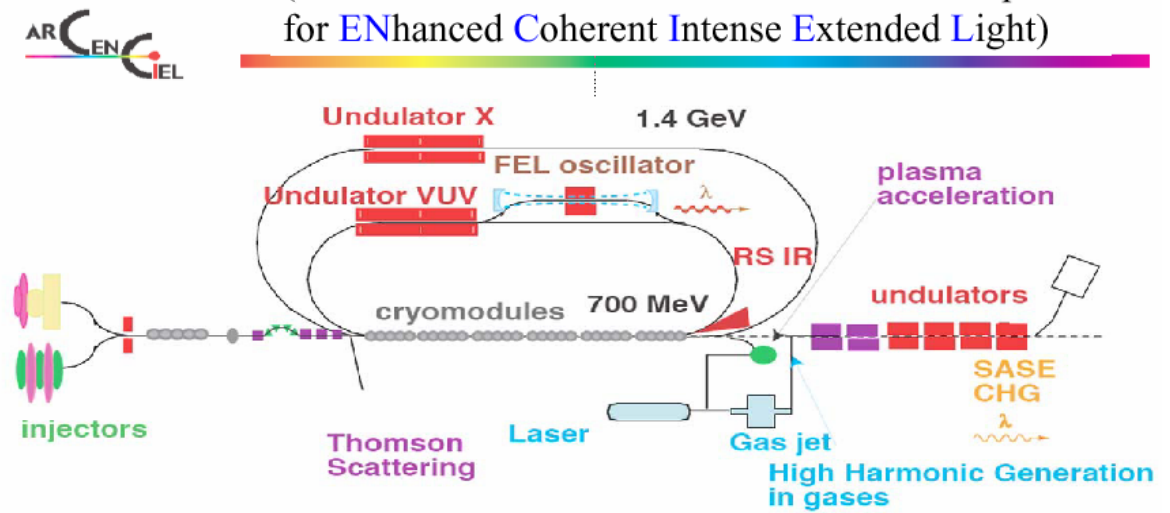

M.E. Couprie et al, "*ARC-EN-CIEL*" *A proposal for a 4th generation light source in France*, Proc. EPAC 2004, 366.

Figure 3

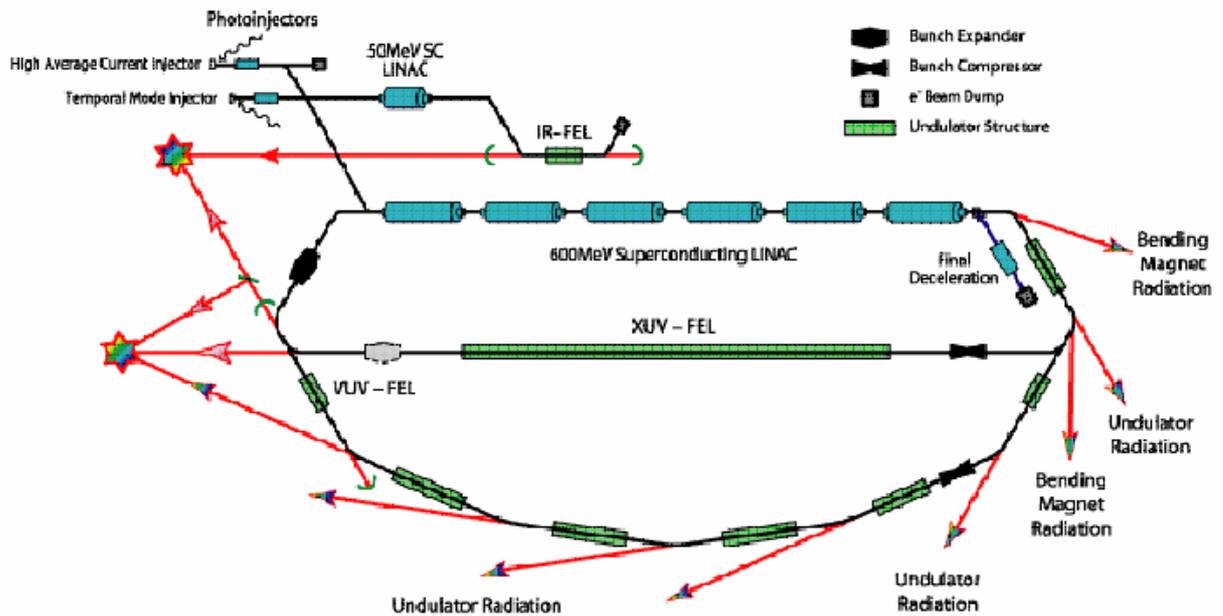

Figure 4

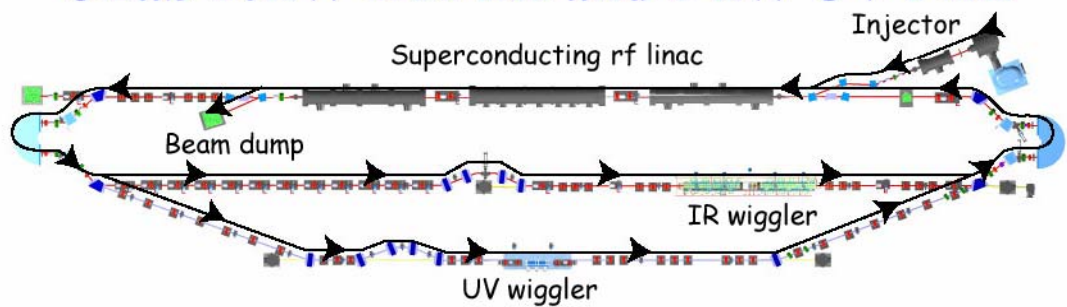

Figure 5
| Output Light Parameters | IR | UV |
|---|---|---|
| Wavelength range (microns) | 1.5 - 14 | 0.25 - 1 |
| Bunch Length (FWHM psec) | 0.2 - 2 | 0.2 - 2 |
| Laser power / pulse (microJoules) | 100 - 300 | 25 |
| Laser power (kW) | >10 | > 1 |
| Rep. Rate (cw operation, MHz) | 4.7 – 75 | 4.7 – 75 |

| Electron Beam Parameters | IR | UV |
|---|---|---|
| Energy (MeV) | 80-200 | 200 |
| Accelerator frequency (MHz) | 1500 | 1500 |
| Charge per bunch (pC) | 135 | 135 |
| Average current (mA) | 10 | 5 |
| Peak Current (A) | 270 | 270 |
| Beam Power (kW) | 2000 | 1000 |
| Energy Spread (%) | 0.50 | 0.13 |
| Normalized emittance (mm-mrad) | <30 | <11 |
| Induced energy spread (full) | 10% | 5% |

S. Benson et al, *High power lasing in the IR upgrade at Jefferson Lab*, 2004 FEL

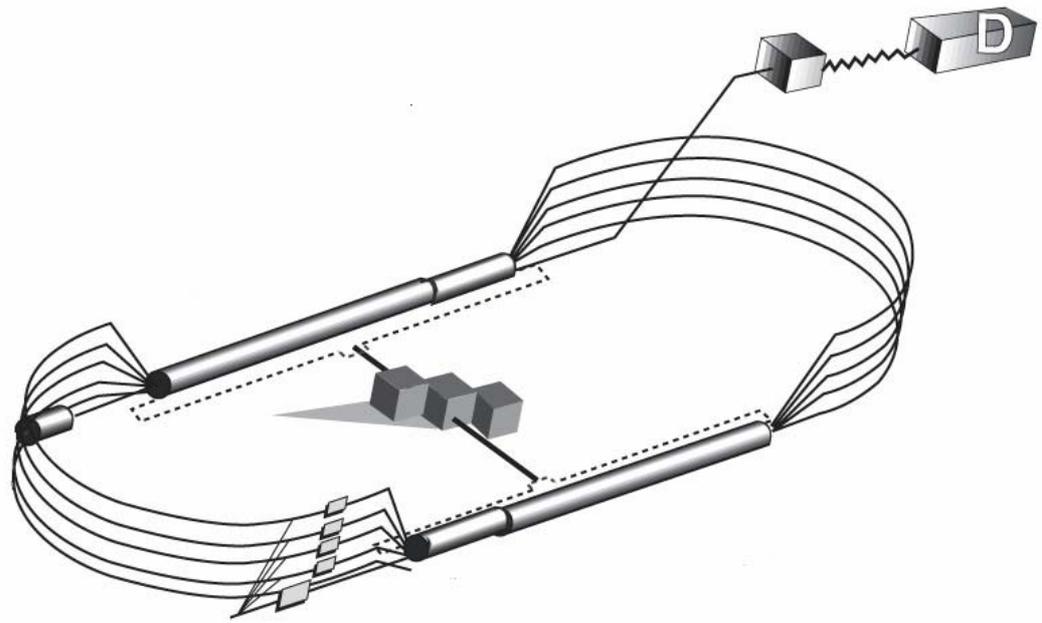

Figure 6: Schematic CEBAF orbits layout.

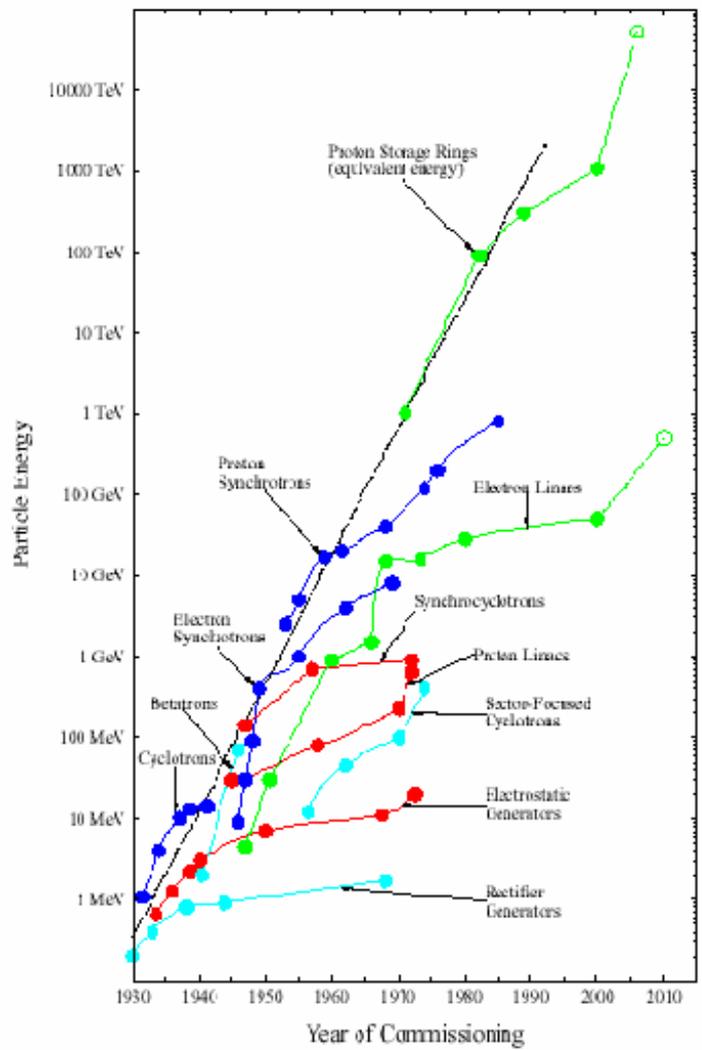

Figure 7.